\documentclass[reprint,twocolumn,bibnotes,amsmath,amssymb,aps,prb,showpacs,floatfix,superscriptaddress,longbibliography]{revtex4-2}
\usepackage[T1]{fontenc}
\usepackage[breaklinks=true,colorlinks,citecolor=blue,linkcolor=blue,urlcolor=blue]{hyperref}
\usepackage{epsfig,mathrsfs,color,latexsym,subfigure,marginnote,gensymb}
\usepackage{graphicx}
\usepackage{multirow}
\usepackage{mathtools}
\usepackage{gensymb}
\renewcommand{\BibitemShut}[1]{}
\usepackage{siunitx}
\usepackage{amsmath}

\begin{document}
\author{Akshay Mahajan}%
\email{amahajan@iitk.ac.in}%
\affiliation{Department of Materials Science and Engineering, Indian Institute of Technology, Kanpur, Kanpur 208016, India}
\author{Somnath Bhowmick}%
\email{bsomnath@iitk.ac.in}%
\affiliation{Department of Materials Science and Engineering, Indian Institute of Technology, Kanpur, Kanpur 208016, India}

\title{Prediction of Magnetoelectric Multiferroic Janus Monolayers VOXY(X/Y = F, Cl, Br, or I, and X$\not=$Y) with in-plane ferroelectricity and out-of-plane piezoelectricity}



\date{\today}
\begin{abstract}
Multifunctional two-dimensional (2D) multiferroics are promising materials for designing low-dimensional multipurpose devices. The key to multifunctionality in these materials is breaking the space-inversion and the time-reversal symmetry, which results in spontaneous electric polarization and magnetization in the same phase. A new class of 2D materials, Janus 2D materials, has emerged, which works on a similar principle of breaking out-of-plane symmetry to invoke new exciting functionalities in the 2D materials, such as an out-of-plane piezoelectric polarization. In this work, a new group of 2D multiferroic Janus monolayers VOXY (X/Y = F, Cl, Br, or I, and X$\not=$Y) is derived by breaking the out-of-plane symmetry in the parent multiferroics VOX$_2$ (X = F, Cl, Br, or I). The structural, magnetic, and ferroelectric properties of multiferroics VOX$_2$ are compared with their Janus derivatives. We calculated in-plane and out-of-plane piezoelectric polarization for VOX$_2$ and VOXY series, where VOFCl, VOFBr, VOFI, and VOClI are found to have significant out-of-plane piezoelectric polarization. Our theoretical work predicts a new series of 2D multiferroic materials and encourages their further investigation for high-performance nanoelectronic devices.  
\end{abstract}
\maketitle

\section{Introduction}
Multiferroics are multifunctional materials that exhibit more than one ferroic ordering, such as ferroelectricity, ferromagnetism, ferroelasticity, or ferrotoroidicity, in the same phase. Different combinations of these ferroic orderings can lead to various multiferroic types, like ferroelectric-ferroelastic or ferromagnetic-ferroelastic. Among these different types, the most relevant ones for microelectronics are magnetoelectric multiferroics\cite{Spaldin2019,Hill2000}, which combine ferroelectric and magnetic behaviors. Nowadays, the multiferroic term most often refers only to these magnetoelectric multiferroics because of their technological significance. These multiferroics combine spontaneous electric polarization and magnetization that an applied electric and magnetic field can switch, along with some coupling between the two ferroics in certain instances. 

Depending on the microscopic source of ferroelectricity, multiferroics can be further classified as type-I and type-II multiferroics\cite{Khomskii2009}. Multiferroics with independent sources for ferroelectricity and magnetism belong to type-I, which have large polarization values and high transition temperatures but weak magnetoelectric coupling. Multiferroics in which ferroelectricity originates due to the breaking of space-inversion symmetry by the magnetic ordering belong to type-II, which exhibits strong magnetoelectric coupling but smaller polarization values and low transition temperatures. Type-II multiferroics are rare compared to type-I multiferroics but are more attractive due to the possibility of electric-field control of magnetism, a revolution for low-energy data storage\cite{Chu2008, Song2017}.

Scientists are challenged to discover and engineer atomically thick multiferroics to incorporate multifunctionality for energy-conserving nanoscale devices with entirely new architectures. However, conventional thin films of three-dimensional (3D) multiferroics suffer from quantum tunneling, electron screening, dangling bonds, surface effects, and depolarization\cite{Tang2019,Fong2004,Junquera2003,Dawber2005}. Therefore, new low-dimensional multiferroic materials are vital for advancing nanoelectronics. In recent years, two-dimensional (2D) multiferroics\cite{Tang2019, Gao2021}, because of their clean surfaces and high dielectric constants, have emerged as a viable option to realize multifunctional nanoscale devices.

Another new family of 2D materials that has attracted increasing interest in the 2D research community is the Janus 2D materials\cite{Zhang2020,Li2018}. Since the landmark experimental discovery of graphene in 2004\cite{Novoselov2004}, several 2D materials have been discovered and investigated for their unique properties and promising applications in diverse fields\cite{Khan2020, Ahn2020, Luo2016, Sahoo2016, Pradeep2019}. With the further exploration of the remarkable properties of 2D materials by manipulating their structures using several approaches like asymmetrical functionalization by various molecular groups\cite{Zhang2013,Bissett2014,Sun2016}, the class of Janus 2D materials has emerged from the concept of asymmetrical facial properties\cite{Montes-Garcia2022,Zhang2020}. The breaking of the out-of-plane symmetry in Janus 2D materials arouses many interesting physical phenomena such as strong Rashba spin splitting\cite{Hu2018,Wang2019} and out-of-plane piezoelectric polarization\cite{Dong2017,Guo2017,Zhang2019}. Following the recent experimental investigations where Janus MoSSe monolayer has been fabricated using a modified CVD method\cite{Lu2017,Zhang2017}, several Janus 2D materials have been discovered and investigated both theoretically\cite{Kandemir2018, Peng2019, Guo2019, Ahammed2020, Jiang2021} and experimentally\cite{Hajra2020, Trivedi2020}.
  
Though the first-ever experimental realization of a 2D multiferroic has been confirmed recently in 2022 only for monolayer NiI$_2$\cite{Song2022}, several 2D multiferroics have been discovered in the past decade using density functional theory (DFT) based theoretical methods\cite{Chakraborty2013, Wu2016, Luo2017, Li2017, Huang2018, Zhang2018, Xu2020, Feng2020, Xu2021, Gao2021}. Most of them have been identified as type-I multiferroics with very few type-II 2D multiferroics\cite{Chakraborty2013,Zhang2018}. Recently, several studies have discovered and investigated a new family of 2D type-I multiferroic VOX$_2$ (X = F, Cl, Br, or I)\cite{Ai2019,Tan2019,Mahajan2021,Xu2020a,You2020,Ding2020,Zhang2021a}. VOX$_2$ monolayers show in-plane ferroelectricity while violating the \textit{d}$^0$ rule\cite{Tan2019} since distortion of V ion with partially occupied \textit{d}$^1$ electronic configuration is responsible for ferroelectricity. Among the four members, VOF$_2$ and VOI$_2$ have been predicted to have ferromagnetic (FM) ground state spin ordering\cite{Tan2019,You2020}. Although, later studies predict non-collinear spin texture as the ground state for VOI$_2$ polar structure because the strong spin-orbit coupling (SOC) of the heavy element iodine leads to large Dzyaloshinki-Moriya interaction (DMI)\cite{Xu2020a,Ding2020}. To the best of our knowledge, among all the VOX$_2$ monolayers, only VOCl$_2$ and VOBr$_2$ have experimental data for their 3D counterpart from which VOCl$_2$ and VOBr$_2$ monolayers can be exfoliated\cite{Hillebrecht1997,Mounet2018}. 

We revisited the VOX$_2$ family in the present work and derived the Janus multiferroic VOXY (X/Y = F, Cl, Br, or I, and X$\not=$Y) by creating asymmetry along the out-of-plane direction and breaking the out-of-plane mirror symmetry. We compared the crystal structure of the Janus VOXY monolayers and discovered a significant change in the bond angles and bond lengths from their parent VOX$_2$, especially in the out-of-plane direction. The VOXY monolayers show FM ground-state magnetic ordering, except VOClBr, without considering SOC effects. Electronic band structure calculations reveal VOClI and VOBrI as direct bandgap semiconductors. Considering collinear magnetism, all FM VOXY monolayers, except VOBrI, predicted to have higher Curie temperature than FM VOF$_2$. Magnetocrystalline anisotropy energies (MAEs) for VOXY monolayers were compared to VOX$_2$ monolayers, suggesting a significant possibility for FM collinear magnetism in VOFCl. Ferroelectric polarization values and energy barriers for polarization switching pathways for VOXY monolayers are also compared with the VOX$_2$ series. Out-of-plane electric polarization is also calculated for Janus VOXY monolayers. In the end, piezoelectric coefficients are calculated for VOXY and VOX$_2$ monolayers. Large values for in-plane piezoelectric coefficients (\textit{e$_{11}$} from 11.12 to 16.90 $\times 10^{-10}$ C/m, \textit{e$_{12}$} from -2.54 to -3.91 $\times 10^{-10}$ C/m) are obtained for all the VOXY monolayers which is also observed in their parent VOX$_2$ monolayers as well. Out-of-plane piezoelectric polarization values for VOFCl, VOFBr, VOFI, and VOClI are comparable to other 2D Janus materials. Our work introduces new series of 2D Janus multiferroics and encourages their further theoretical and experimental investigations for new advanced architectures in nanodevices.

\begin{figure}[!htbp]
\includegraphics[width=\linewidth]{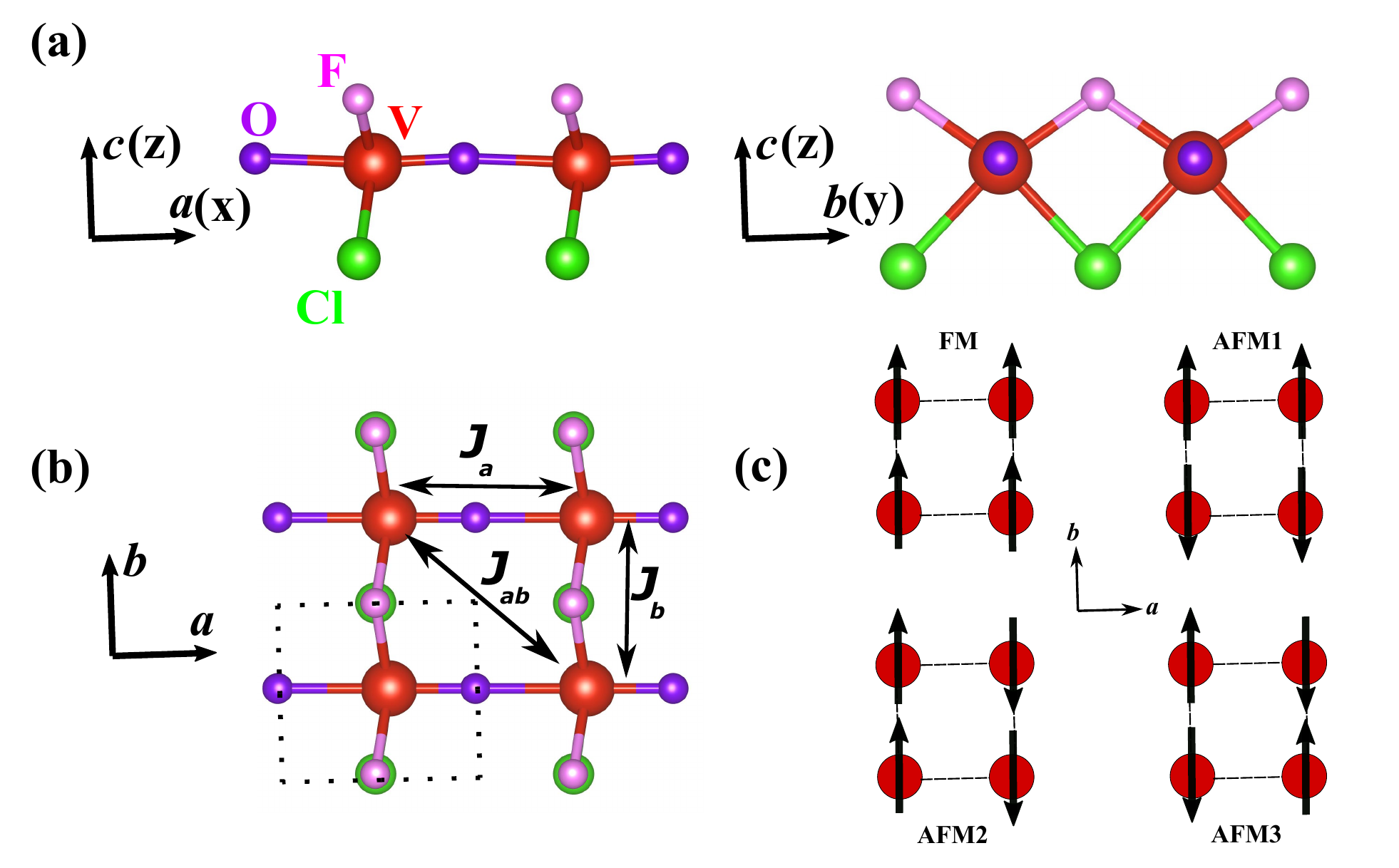}
\caption{(a) Side and (b) Top views of the $2\times2\times1$ supercell of VOFCl monolayer. Note that \textit{a}, \textit{b}, and \textit{c} lattice parameters are along the \textit{x-}, \textit{y-}, and \textit{z-}direction, respectively. Vacuum is along the \textit{c-}axis, with \textit{a} and \textit{b} as in-plane lattice parameters. Black dotted rectangle in (b) encloses the primitive cell. The magnetic exchange coupling parameters (\textit{J$_a$}, \textit{J$_b$}, and \textit{J$_{ab}$}) are also shown in (b). (c) Four different magnetic configurations considered in this work. Only V ions are shown in the schematic [compare with (b)]}       
\label{f1}
\end{figure}
            
\section{Computational Details}
\label{secdetails}
Our density functional theory (DFT) calculations were performed using the Vienna \textit{ab initio} Simulation Package (VASP) suite of codes.\cite{Kohn1965,Kresse1996,Kresse1996PRB} The projector augmented wave (PAW)\cite{Blochl1994,Kresse1999} pseudopotentials were used in which the generalized gradient approximation (GGA) formulation of Perdew-Burke-Ernzerhof (PBE) \cite{Perdew1996} accounts for the exchange and correlation effects. Excluding phonon spectra calculations, for all DFT calculations dispersion interactions were taken into account using vdW-DF2\cite{Dion2004,Roman2009,Lee2010,Klime2011}. A $2\times2\times1$ supercell was constructed to consider different possible magnetic orderings. An energy cutoff of 550 eV was used with a gamma-centered \textit{k}-grid of $9\times9\times1$ to sample the Brillouin zone (BZ). During structural relaxation, atomic positions and lattice constants are optimized using a conjugate gradient (CG) algorithm with the criterion of 0.001 eV/{\AA} for the Hellman-Feynman forces. The energy convergence criterion was fixed to 10$^{-6}$ eV. A vacuum layer of 20 {\AA} was added in the out-of-plane direction to avoid interactions between the monolayer and its periodic images. The LDIPOLE tag in VASP was used to apply dipole correction for correcting the errors introduced by the periodic boundary conditions.\cite{Neugebauer1992} The phonon spectra calculations were done using DFPT+ Phonopy method.\cite{Gonze1997, phonopy} Magnetic Anisotropy Energies (MAEs) for each monolayer were calculated using the magnetic force theorem.\cite{Izardar2020,LIECHTENSTEIN1987,Daalderop1991} During spin-orbit coupling (SOC) calculations to determine MAE, a denser \textit{k}-grid of $11\times11\times1$ was used for better convergence of MAE values. The electric polarization values were computed using the standard Berry phase method.\cite{King1993,Resta1994} See the additional note on electric polarization calculations performed in this work in the Supplemental Material\cite{[URL will be inserted by publisher]}. For the calculation of the piezoelectric coefficients, strain is defined as $\varepsilon$ = \textit{(a-a$_o$)/a$_o$} = \textit{(b-b$_o$)/b$_o$}, where \textit{a}, \textit{b} are in-plane lattice parameters after applied strain and \textit{a$_o$}, \textit{b$_o$} are the equilibrium values for the in-plane lattice parameters.

\section{Results and Discussions}
\label{secresults}
In order to create Janus VOXY monolayers, we started from the ground-state structures of parent VOX$_2$ monolayers and replaced the top layer of halogen atoms with another halogen of higher electronegativity. The higher electronegative halogen atom in the upper layer results in an electric dipole moment within the monolayer pointing towards the positive \textit{c-}axis. Figure \ref{f1} shows the ground-state structure of the VOFCl monolayer, which is the template for the atomic structure of the other VOXY monolayers. The VOXY structures have lower symmetry \textit{Pm} compared to their parent VOX$_2$ monolayers, which belong to a space group of \textit{Pmm2}. Lowering the symmetry in the VOXY monolayers results from breaking the out-of-plane mirror symmetry. Similar to their parent VOX$_2$ monolayers, V ion displacement from its in-plane centrosymmetric position along the \textit{a-}axis persist in the VOXY monolayers, and thus they are supposed to show similar in-plane ferroelectric properties\cite{Ai2019,Tan2019,You2020,Mahajan2021}. We then verified the stability of these Janus VOXY monolayers by obtaining their phonon spectra, as shown in Figure S1 of the Supplemental Material\cite{[URL will be inserted by publisher]}. Since no significant imaginary phonon frequency is observed for any VOXY monolayers, we believe these monolayers to be dynamically stable.     
 
The structural parameters for VOXY monolayers are compared with the parent VOX$_2$ monolayers in Table S1 of the Supplemental Materials\cite{[URL will be inserted by publisher]}. The lattice constant \textit{b} increases with the addition of a heavier halogen atom in the monolayer while lattice parameter \textit{a} does not show significant variation for both VOXY and VOX$_2$ monolayers, similar to the previous works on the VOX$_2$ monolayers\cite{Tan2019}. The \textit{b} lattice constant for VOXY monolayers has values between those for their parent VOX$_2$ monolayers, which is expected. The thickness of the monolayers, which can be considered as the distance between halogen layers as given by X-Y length in Table S1\cite{[URL will be inserted by publisher]}, increases with the increase in the radii of the halogen atoms. The vanadium-halogen bond length, given as V-X/Y in Table S1 where X is halogen with higher electronegativity, decreases with an increase in the electronegativity of the halogen atom, as can be seen in Table S1 where V-X is always smaller than V-Y. The vanadium-halogen bond length also depends on the monolayer's other halogen. For example, comparing the V-X(=F) bond lengths in VOFCl, VOFBr, and VOFI, it increases with the addition of lower electronegative Y halogen atoms (Cl, Br, and I) as 2.028 {\AA}, 2.043 {\AA}, and 2.067 {\AA}, respectively, which are higher than the V-F bond length in VOF$_2$ of 2.001 {\AA}. Similarly, the V-Cl bond increases in VOClBr to 2.597 {\AA} compared to 2.435 {\AA} in VOCl$_2$. The difference in vanadium-oxygen bond lengths (V-O1 and V-O2) in each VOXY monolayer provides the polar displacement of the V ion along the in-plane polar axis (\textit{a-} axis), shown in Table S1. This V ion polar displacement is responsible for possible in-plane ferroelectricity in these systems similar to their parent VOX$_2$ monolayers as investigated in our previous work for VOCl$_2$\cite{Mahajan2021}. X-V-X bond angle is larger than the Y-V-Y bond angle due to higher repulsion among halogen atoms with higher electronegativity. X-V-X bond angle increases further with the decrease in electronegativity of Y halogen and vice versa for the Y-V-Y bond angle, which decreases in the presence of higher electronegative X halogen. Out-of-plane asymmetry in VOXY monolayers also results in buckling in the V-O chain, as seen from the O-V-O bond angle values differing from 180\degree.

We considered four different magnetic orderings, one ferromagnetic (FM) and three antiferromagnetic (AFM1, AFM2, and AFM3), for the VOXY monolayers similar to the ones considered in the previous studies for their parent VOX$_2$ monolayers\cite{Ai2019,Tan2019,You2020,Mahajan2021}, which are shown in Figure \ref{f1}(c). The energy values for these four different magnetic orderings are compared in Table \ref{table1}. All the VOXY monolayers have FM as the lowest energy magnetic ordering, except for VOClBr, which has AFM3 as the minimum energy ordering. From these results, we can predict that for the possibility of an intrinsic ferromagnetic ground state for VOXY monolayers, either F or I is required within the structure. This result is consistent with those for VOX$_2$ monolayers, where only VOF$_2$ and VOI$_2$ monolayers show FM as the lowest energy magnetic ordering. We also calculated the V ion’s magnetic moment for each VOXY and VOX$_2$ monolayers, as shown in Table \ref{table1}. A higher magnetic moment is obtained for monolayers with FM ordering, and the magnetic moment increases with adding a heavier halogen atom to the structure.

\begin{figure}
\includegraphics[width=\linewidth]{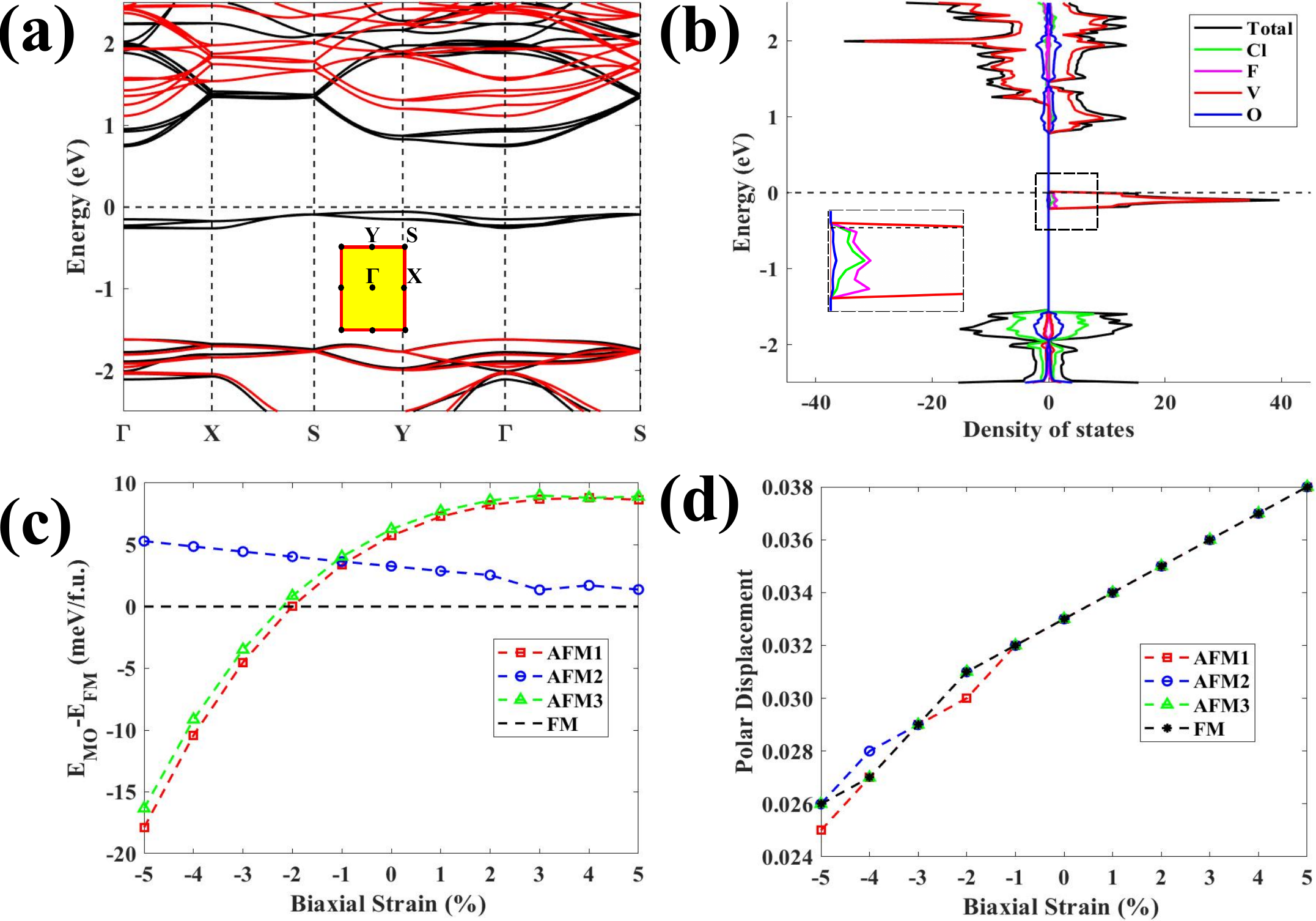}
\caption{(a) Electronic band structure of the VOFCl monolayer. Inset shows the high-symmetry points in the first Brillouin zone. Black and red colour bands represent spin-up and spin-down bands, respectively. (b) Orbital-resolved density of states of the VOFCl monolayer. Inset shows the valence states closest to the Fermi level. (c) Variation of energy difference between different magnetic orderings (MO) and FM magnetic ordering with biaxial strain for the VOFCl monolayer. (d) Change in the V ion's polar displacement with biaxial strain for different magnetic orderings in the VOFCl monolayer.}       
\label{f2}
\end{figure}

Figure \ref{f2} (a) shows the electronic band structure for the VOFCl monolayer with FM magnetic ordering. We observed an indirect band gap of 0.802 eV and a direct band gap of 2.738 eV for the spin-up (black) and spin-down (red) bands, respectively. The Supplemental Material provides electronic band structure plots for other VOXY  and VOX$_2$ monolayers in Figures S2 and S3, respectively\cite{[URL will be inserted by publisher]}. For all VOXY monolayers with FM orderings, one can observe that the spin-down (red) bands have a direct band gap at the $\Gamma$ point larger than the one for spin-up (black) bands. Interestingly, VOClI and VOBrI monolayers show a direct band gap for both spin-up (black) and spin-down (red) bands at the $\Gamma$ point. For VOClI monolayer, the electronic band gaps are 0.423 eV and 0.853 eV for spin-up (black) and spin-down (red) bands, respectively. Furthermore, for the VOBrI monolayer, the electronic band gaps are 0.358 eV and 0.833 eV for spin-up (black) and spin-down (red) bands, respectively. Thus, we can predict VOClI and VOBrI monolayers as 2D direct band gap semiconductors with FM ordering. 
 
Table \ref{table1} shows the electronic band gaps for all the VOXY and VOX$_2$ monolayers. The lowest band gap among the spin-up and spin-down bands should be considered the majority band gap. We observe that heavier halogen atoms lead to the lowering of the electronic band gaps. Also, the VOClBr monolayer has the highest majority band gap of 0.997 eV among the VOXY monolayers, which can be attributed to its antiferromagnetic nature. Figure S3 in the Supplemental Material\cite{[URL will be inserted by publisher]} and band gap data from Table \ref{table1} show that VOF$_2$ has an indirect majority band gap, while VOI$_2$ has a direct majority band gap (although of a minimal magnitude of 0.001 eV). Comparing this with the data for other VOXY monolayers in Table \ref{table1}, we can conclude that while I in VOXY structure can lead to a direct band gap, the presence of F results in an indirect band gap. 

Figure \ref{f2} (b) illustrates the orbital-resolved density of states (DOS) plot for the VOFCl monolayer. We observe that the V atoms predominantly contribute to valence and conduction states near the Fermi level. On further zooming into the valence states in the DOS, as shown in Figure \ref{f2} (b), we observe that there are also minor contributions from the F and Cl atoms to the valence states, and among them, the F atom has a more considerable contribution. Note that no significant contribution from O atoms to the valence states is observed. Figure S4 shows the DOS plots for other VOXY monolayers\cite{}. The predominance of V atom in near Fermi level valence and conduction states with minor contributions from halogen atoms in the valence states remains consistent for all VOXY monolayers. In the case of VOFBr and VOFI monolayers, similar to VOFCl, F atoms have a more significant contribution compared to Br and I atoms, respectively. While in the case of VOClI and VOBrI monolayers, I atoms have a more considerable contribution than Cl and Br atoms, respectively. For the VOClBr monolayer, both Cl and Br atoms show a similar contribution to the valence state with a little higher from the Cl atoms. Combining the results from these DOS plots and the comparison we have done of the properties of VOXY monolayers with VOX$_2$ monolayers, we can conclude that the halogen atom with the higher contribution in the near Fermi level valence state can significantly affect the electronic and magnetic properties of the VOXY monolayers. For example, the VOFI monolayer has an indirect band gap just like the VOF$_2$ monolayer instead of a direct band gap like the VOI$_2$ monolayer since the valence state in the VOFI monolayer has a higher contribution from F atoms rather than I atoms. Similarly, the FM nature of the VOFCl and VOFBr monolayers could be attributed to the higher contribution of the F atoms in the valence states than other halogen atoms in those monolayers. In the case of VOClI and VOBrI monolayers, we can also correlate the direct band gap observed in these monolayers to the higher I atom contribution in the valence states.

In our previous work\cite{Mahajan2021}, we observed that we could tune the ground-state magnetic ordering and the ferroelectric polarization due to the change in the V ion’s polar displacement of the VOCl$_2$ monolayer using strain-engineering. A similar effect of strain-engineering on the magnetic and ferroelectric properties of the VOF$_2$ monolayer is also reported.\cite{You2020} Since Janus monolayers exhibit properties similar to their parent monolayers, we applied biaxial strain on the VOFCl monolayer to determine the effect of strain-engineering on the energies of different magnetic orderings and the V ion’s polar displacement in the VOFCl monolayer, as shown in Figure \ref{f2} (c) and (d), respectively. We observe a change in the lowest energy magnetic ordering from FM to AFM1 at around 2\% compressive strain. Also, with increasing tensile strain, the FM ordering becomes more stable compared to AFM1 and AFM3 since their energy differences increase. Similar results have also been observed for VOF$_2$ monolayer\cite{You2020} and for VOCl$_2$ monolayer\cite{Mahajan2021}, where the ground state magnetic ordering change from AFM3 to FM on applying tensile strain. The V ion’s polar displacement increases with the increase in tensile strain, and vice versa decreases for compressive strain in the VOFCl monolayer, with similar polar displacement values for different magnetic orderings. VOCl$_2$ also shows a similar trend for change in the polar displacement with strain for different magnetic orderings, as reported in our previous work\cite{Mahajan2021}.

\begin{figure}
\includegraphics[width=\linewidth]{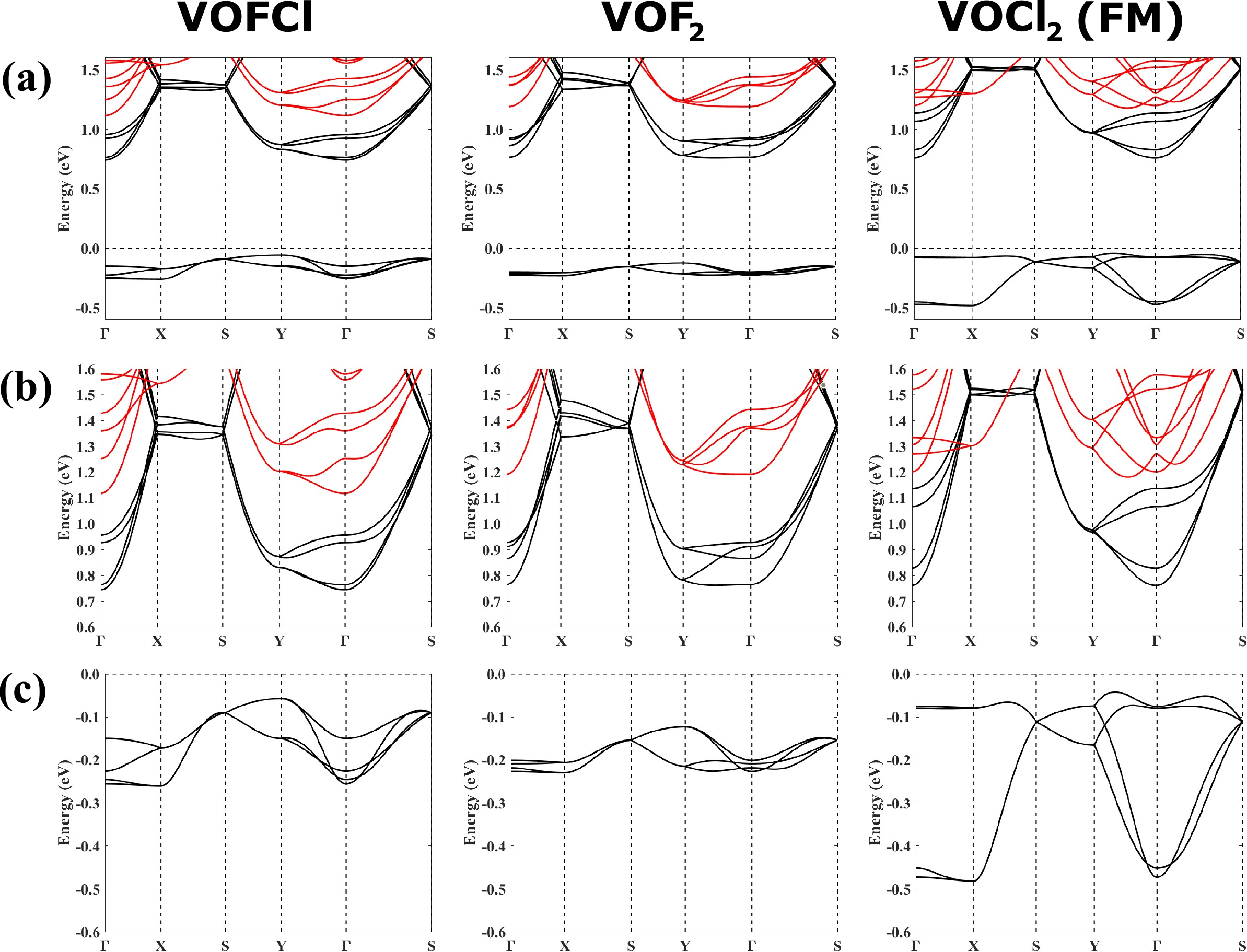}
\caption{Comparison of the (a) electronic band structure near Fermi level, (b) conduction band, and (c) valence band, for the VOFCl, VOF$_2$, and FM ordered VOCl$_2$ monolayers. Black and red colour bands represent spin-up and spin-down bands, respectively. }
\label{f3}
\end{figure}

Continuing with our comparative analysis of the properties of VOXY and VOX$_2$ monolayers, we revisited the band structure of the VOFCl monolayer and compared it with the electronic band structure of the VOF$_2$ monolayer and VOCl$_2$ monolayer, with FM ordering. Figure \ref{f3} (a) shows the near Fermi level band structure for VOFCl, VOF$_2$, and VOCl$_2$ (FM). We can observe that band structures of both VOFCl and VOF$_2$ show similar characteristics with certain distinctive differences, other than the electronic band gap, in both the conduction and valence bands. Figure \ref{f3} (b) shows the enhanced plots for the conduction bands of VOFCl, VOF$_2$, and VOCl$_2$ (FM). In the VOF$_2$ conduction band, we observe two degenerate spin-up (black) bands along the $\Gamma$-Y direction and two degenerate spin-down (red) bands at the $\Gamma$ point. In the case of the VOFCl monolayer’s conduction band, such degeneracies are observed to be lifted. For the VOCl$_2$ (FM) conduction band, similar to VOFCl, we observe four non-degenerate spin-up (black) bands along the $\Gamma$-Y direction but a four-fold degeneracy at the Y point. Figure \ref{f3} (c) shows the enhanced version of the valence band of VOFCl, VOF$_2$, and VOCl$_2$ (FM) monolayers. Both VOFCl and VOF$_2$ have similar bands for the valence states with higher bandwidth in the case of the VOFCl monolayer, which can be attributed to the Cl atoms since the VOCl$_2$ (FM) monolayer also has large bandwidth for the valence band.

Then the magnetic exchange coupling parameters are calculated for nearest-neighbor (NN) V ions along the \textit{a} and \textit{b} axes represented by \textit{J$_a$} and \textit{J$_b$}, respectively, and next-nearest-neighbor (NNN) V ions represented by \textit{J$_{ab}$}, as shown in Figure \ref{f1}(b). Using the Ising model Hamiltonian presented in our previous work\cite{Mahajan2021}, we can write following energy equations for different magnetic orderings -:
\begin{eqnarray}
E_{FM}= E_0 - 4M^2_{FM} (J_a+J_b+2J_{ab}),\\
E_{AFM1} = E_0 - 4M^2_{AFM1} (J_a-J_b-2J_{ab}),\\
E_{AFM2} = E_0 - 4M^2_{AFM2} (-J_a+J_b-2J_{ab}),\\
E_{AFM3} = E_0 - 4M^2_{AFM3} (-J_a-J_b+2J_{ab}).
\end{eqnarray}
Here E$_0$ is the non-magnetic part of the energy, and M$_{MO}$ (where MO=FM, AFM1, AFM2, AFM3) is the magnetic moment for different magnetic orderings. Using the energy values for different magnetic orderings (provided in Table \ref{table1}) in the above equations, we can extract the values for exchange parameters, \textit{J$_a$}, \textit{J$_b$}, and \textit{J$_{ab}$}, for all the VOXY and VOX$_2$ monolayers.  

The calculated exchange parameters are then used to provide an estimate for the transition temperature within the mean-field approximation\cite{Strecka2015} using the formula below-: 
\begin{equation}
\begin{split}
T & = \frac{M^2}{k_B} |2J_a + 2J_b + 4J_{ab}|
\end{split}
\label{TCEQUATION}
\end{equation}        
Here, k$_B$ is the Boltzman constant, and M is the magnitude of the V ion’s magnetic moment for the ground-state spin-order. 

The calculated values for exchange parameters and the transition temperatures (T$_C$ and T$_N$) for VOXY and VOX$_2$ monolayers are provided in Table \ref{table2}. Note that the negative (positive) exchange parameter suggests anti-parallel (parallel) ordering between the coupled pair, and the values of exchange parameters affect the magnetic ordering as per the equations discussed earlier. Thus, for VOXY and VOX$_2$ monolayers with all exchange parameters positive, V ion spins are parallelly ordered to each other, resulting in an FM ground state order. For VOBrI, although \textit{J$_b$} is negative, it is closer to zero and has a much lower value than other positive exchange parameters resulting in FM order. For VOClBr, VOCl$_2$ and VOBr$_2$, the high negative value for \textit{J$_b$} results in anti-parallel spin-ordering along the \textit{b-}axis, and the  \textit{J$_a$} value is smaller than the twice of the \textit{J$_{ab}$} value resulting parallel ordering along the direction of \textit{J$_{ab}$} coupling resulting in AFM3 order.

By comparing the T$_C$ (Curie Temperature) values for FM monolayers, we found that for the monolayers with larger radii I atom, the T$_C$ value decreases with the increase in the radii of the other halogen atom. On the other hand, for the FM monolayers with F atoms, adding a larger radii halogen increases the T$_C$ value resulting in a higher T$_C$ value for VOFBr compared to VOFCl, which is larger than that for VOF$_2$ as well. For AFM3 monolayers, larger halogen radii reduce T$_N$ (Neel Temperature) values, resulting in lower T$_N$ value for VOClBr than VOCl$_2$. Although the calculated transition temperature (T$_C$ and T$_N$) values are expected to be overestimated, comparing the values gives a qualitative understanding of magnetism in the VOXY and VOX$_2$ monolayers in the case of a collinear magnetic ground state

Magnetic anisotropy is an essential requirement for stabilizing long-range magnetic ordering in the 2D materials by lifting the restrictions imposed by the Mermin-Wagner theory\cite{Mermin1966,Huang2017}. To determine the stability of the magnetic orderings in the VOXY monolayers, we calculated the magnetic anisotropy energies (MAEs) by considering the spin-orbit coupling (SOC) approximation. The values for MAEs for all the VOXY and VOX$_2$ monolayers are given in Table \ref{table3}. For MAE calculations, we considered magnetization along the [100], [010], [001], [110], and [111] directions. We observe that VOFCl, VOBrI, VOF$_2$, and VOI$_2$ monolayers have a magnetic easy axis along the [010] direction or the \textit{b}-axis, which is in-plane but perpendicular to the polar axis, \textit{a}-axis. On the other hand, for VOFI and VOClI monolayers, the easy axis is along the polar \textit{a}-axis, [100] direction. For VOClBr, VOFBr, VOCl$_2$, and VOBr$_2$ monolayers, the magnetic easy axis points out-of-plane along the \textit{c}-axis, [001] direction.
 
In some recent studies\cite{Xu2020a,Ding2020}, a spiral spin texture has been reported as the ground state for VOI$_2$ monolayer instead of a collinear FM ordering. This distortion of FM spin texture of VOI$_2$ monolayer to a non-collinear ground-state has been pointed out as the result of the strong Dzyaloshinskii-Moriya interaction (DMI)\cite{Dzyaloshinsky1958,Moriya1960}. Although DMI is directly associated with the polar distortion in the VOX$_2$ monolayers, it was revealed that the presence of heavy element I, which provides a strong spin-orbit coupling(SOC), results in an effective DMI in the polar structure that can lead to the distortion of collinear FM ordering to a short-period spiral structure. On the other hand, VOF$_2$ has been reported to have a collinear FM ground state since it shows weak magnetoelectric anisotropy and DMI in the absence of any heavy element with a strong SOC like iodine\cite{You2020}. For FM VOXY monolayers, we can compare the MAE values with those of VOF$_2$ and VOI$_2$ monolayers from Table \ref{table3} and observe that VOFCl shows smaller values compared to even VOF$_2$ monolayer, while VOFBr, VOFI, VOClI, and VOBrI shows MAE values much larger than the ones for VOF$_2$ but still smaller than those for VOI$_2$ monolayer. Although properly commenting on the actual magnetic ground states for these VOXY monolayers even within DFT approximation requires some advanced calculations, from our results and conclusions of previous studies, we can predict that VOFCl has an excellent possibility of having a collinear FM ground state. In addition, VOFCl is expected to have a higher T$_C$ value for collinear FM ordering than VOF$_2$ monolayer, as shown in Table \ref{table2}, making it a desirable candidate for further exploration for a 2D collinear ferromagnet.

\begin{figure}
\includegraphics[width=\linewidth]{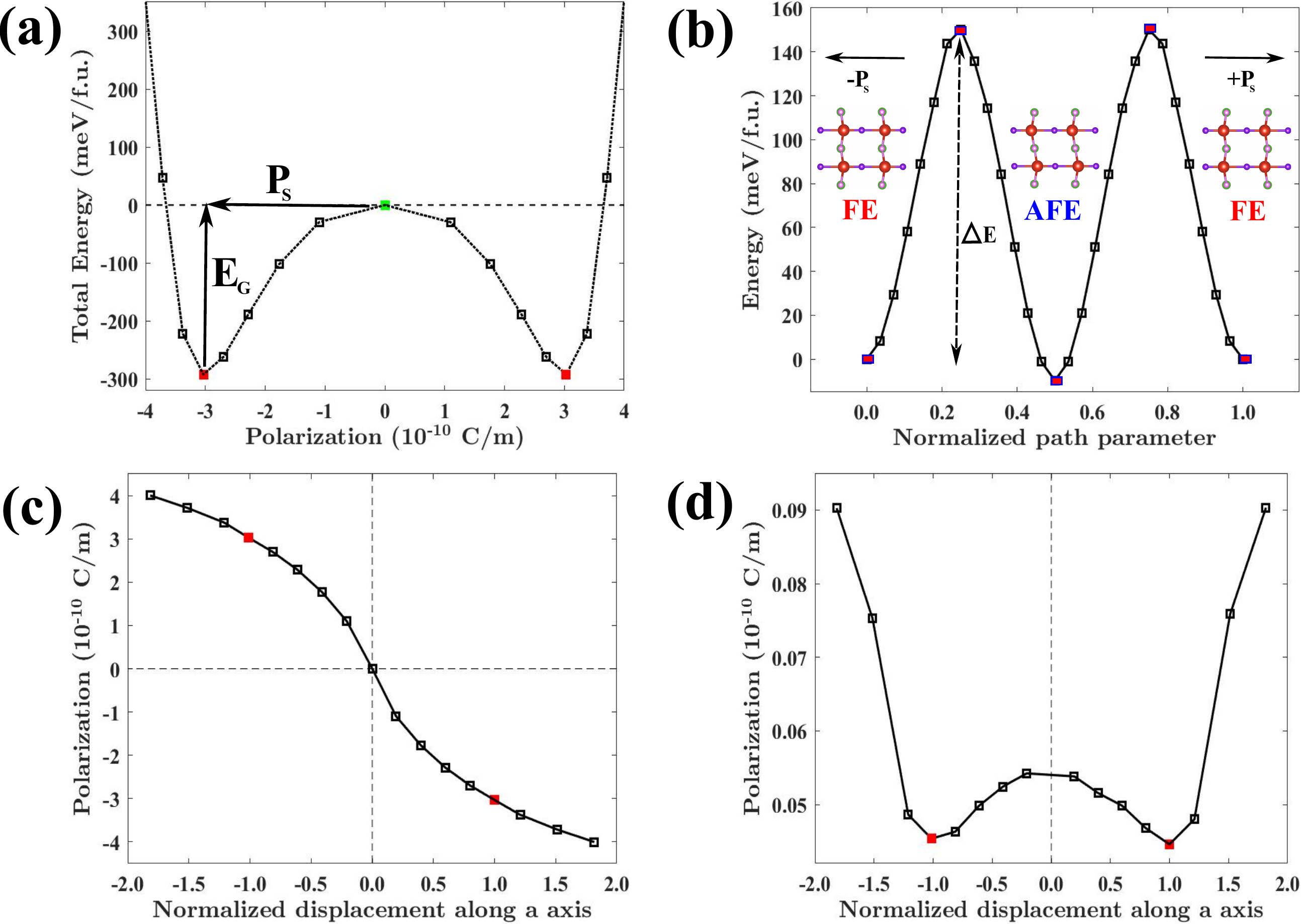}
\caption{In-plane polarization switching pathway in the VOFCl monolayer via (a) paraelectric (PE) inetrmediate phase (shown as green square) with energy barrier E$_G$, and (b) antiferroelectric (AFE) intermediate phase with energy barrier $\Delta$E. P$_S$ corresponds to the spontaneous electric polarization of the ferroelectric ground state. Calculated (c) in-plane and (d) out-of-plane polarization as a function of V ion's displacement along the polar \textit{a}-axis for VOFCl monolayer. Red squares in (a), (c) and (d) corresponds to ferroelectric ground state monolayers.}
\label{f4}
\end{figure}

Next, we analyze the in-plane ferroelectric properties and the out-of-plane polarization in the VOXY monolayers. Figure \ref{f4} (a) shows the energy versus in-plane polarization for the VOFCl monolayer, which looks like a double-well potential curve, a characteristic feature of ferroelectrics with switchable spontaneous electric polarization. VOFCl has a spontaneous in-plane ferroelectric polarization (P$_S$) of 3.03 $\times$ 10$^{-10}$ C/m with an energy barrier (E$_G$) of 292.11 meV/f.u.(formula unit) for polarization switching via a high-symmetry paraelectric phase. Previous work on VOCl$_2$ monolayer revealed a polarization switching pathway via an antiferroelectric (AFE) phase as the pathway with the lowest energy barrier according to the climbing image nudged elastic band calculations\cite{Ai2019}. Figure \ref{f4} (b) shows a similar in-plane ferroelectric polarization switching pathway via the AFE phase for the VOFCl monolayer. The energy barrier for polarization switching via the FE-AFE-FE pathway is defined by $\Delta$E, as shown in Figure \ref{f4} (b), which is equal to 150.22 meV/f.u. for VOFCl, smaller than that of E$_G$. Note that the AFE phase in Figure \ref{f4} (b) has lower energy than the VOFCl monolayer's FE phase. Figure \ref{f4} (c) shows the variation of in-plane polarization with the displacement of V ion along the polar axis (\textit{a}-axis). We can observe that the in-plane polarization switches direction from being along the negative \textit{a}-axis to the positive \textit{a}-axis with the V ion’s displacement from its centrosymmetric position to along the positive and negative \textit{a}-axis, respectively. Figure \ref{f4} (d) shows the variation of out-of-plane polarization with the V ion’s displacement along the \textit{a}-axis. Although the V ion’s displacement affects the magnitude of out-of-plane polarization in such a way that it forms a double-well curve, the polarization does not change its direction, which was expected.
 
Table \ref{table4} provides the values for spontaneous in-plane ferroelectric polarization (P$_S$), the energy barriers for polarization switching (E$_G$ and $\Delta$E), and the energy difference between FE and AFE phase, and the out-of-plane polarization for the VOXY and VOX$_2$ monolayers. Among the VOXY monolayers, VOFCl and VOBrI show the highest and the lowest in-plane ferroelectric polarization of 3.03 $\times$ 10$^{-10}$ C/m and 2.37 $\times$ 10$^{-10}$ C/m, respectively. From the obtained data for P$_S$, we can conclude that the heavier halogen atoms lead to a decrease in P$_S$ values in the VOXY monolayers, which is supported by the results for VOX$_2$ monolayers where VOF$_2$ with lighter F has highest P$_S$ value of 3.31 $\times$ 10$^{-10}$ C/m while VOI$_2$ with heavier I has the lowest P$_S$ value of 2.13 $\times$ 10$^{-10}$ C/m. The energy barriers, E$_G$ and $\Delta$E, follow a similar trend as the P$_S$ values with the highest E$_G$ and $\Delta$E values of 292.11 meV/f.u. and 150.22 meV/f.u., respectively, for VOFCl monolayer, and lowest E$_G$ and $\Delta$E values of 145.51 meV/f.u. and 71.12 meV/f.u., respectively, for VOBrI monolayer, among the VOXY monolayers. This direct relation between ferroelectric polarization and polarization switching energy barrier is expected, in general, for ferroelectric materials. We can observe that $\Delta$E values are always smaller than the E$_G$ values for all the VOXY and VOX$_2$ monolayers, and thus we can conclude that the FE-AFE-FE pathway, shown in Figure \ref{f4} (b), is the minimum energy pathway for in-plane ferroelectric polarization switching.
 
From Table \ref{table4}, we can also observe that VOFCl, VOFBr, VOF$_2$, and VOI$_2$ monolayers have lower energy for the AFE phase, which suggests an AFE ground state for these monolayers. Note that for Figure \ref{f4} (b) and for the values outside the brackets in the FE$-$AFE column of Table \ref{table4}, the AFE phase has the same lattice parameters as the FE phase. On allowing lattice parameter optimization of the AFE phase, the stability of the AFE phase to the FE phase increases, as can be seen from the values provided in the brackets in the FE$-$AFE column. Although AFE phase has lower energy compared to FE phase in these monolayers, large energy barriers $\Delta$E between FE and AFE phases, which are larger than the room temperature thermal energy of 25 meV, ensure stability of the FE phase. In the end, we also calculated the out-of-plane polarization for the VOXY monolayers and obtained non-zero polarization for all the monolayers with the highest value of 0.11 $\times$ 10$^{-10}$ C/m for the VOFI monolayer, which has highest electronegativity difference between the halogens.

\begin{figure}
\includegraphics[width=\linewidth]{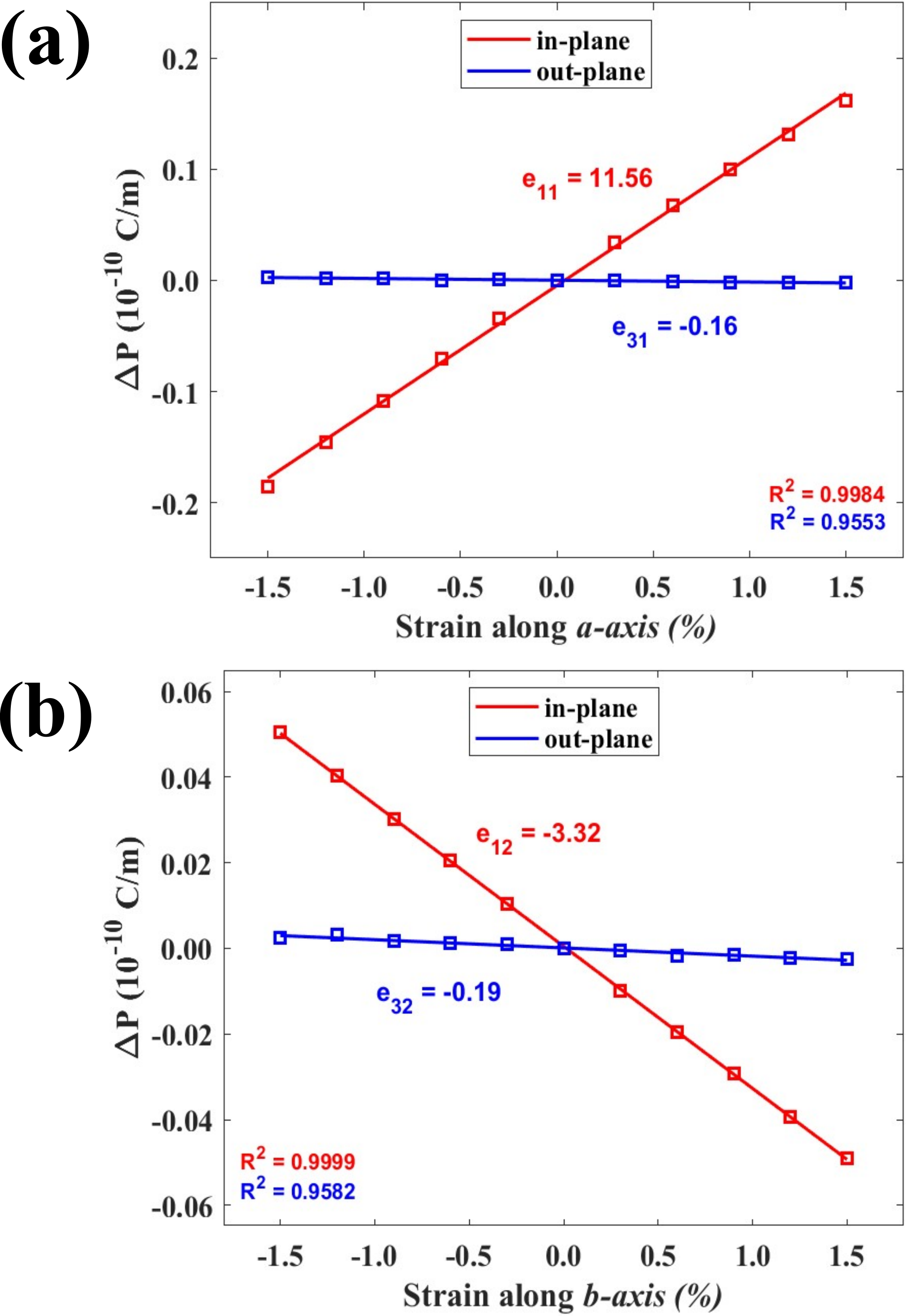}
\caption{Change in the electric polarization for strain (a) along the a-axis and (b) along the b-axis for the VOFCl monolayer. The slope of the linear fit to the curve provides the monolayer's in-plane (e$_{11}$ and e$_{12}$) and out-of-plane (e$_{31}$ and e$_{32}$) piezoelectric coefficients. The R-squared values for the linear fit for both in-plane and out-of-plane piezoelectric polarization are provided in the bottom-right and bottom-left corners in (a) and (b), respectively.}
\label{f5}
\end{figure}

One of the exciting properties that Janus monolayers show is the out-of-plane piezoelectric effect\cite{Zhang2020}. To complete our analysis for the Janus VOXY monolayers, we calculated the in-plane and out-of-plane piezoelectric coefficients for VOXY monolayers and compared their values with those for VOX$_2$ monolayers. The piezoelectric coefficient ($e_{ij}$) is defined as $\frac{\partial P_i}{\partial \epsilon_j}$ which provides the electromechanical coupling between strain and electrical polarization in terms of change in polarization ($\partial P_i$) in the \textit{i}$^{th}$ direction by strain ($\partial \epsilon_j$) in the \textit{j}$^{th}$ direction. Figure \ref{f5} shows the curve for change in the in-plane and out-of-plane polarization with uniaxial strain along \textit{a}-axis and \textit{b}-axis for the VOFCl monolayer. To determine the piezoelectric coefficients, we did a linear fit, using the linear regression method as done in MATLAB, for the data of the magnitude of change in polarization with strain from -1.5\% to 1.5\% in steps of 0.3. The slopes of the obtained linear curves provide the piezoelectric coefficients, as given in Figure \ref{f5}. For VOXY monolayers other than VOFCl, the plots for determining the piezoelectric coefficients are given in Figures S5-S9 in the Supplemental Material\cite{[URL will be inserted by publisher]}.
 
Table \ref{table5} contains the calculated values of the piezoelectric coefficients for all the VOXY and VOX$_2$ monolayers. The \textit{R$^2$} value provides information about the goodness of our linear fit, where a high value of \textit{R$^2$} is considered a better fit with \textit{R$^2$} equal to 1 as the exact fit. For the VOFCl monolayer and other VOXY monolayers, the \textit{R$^2$} values are provided at the bottom of the plots in Figure \ref{f5} and the Figures S5-S9 in the Supplemental Material\cite{[URL will be inserted by publisher]}, respectively. In Table \ref{table5}, we have provided the piezoelectric coefficient values for only those plots with a \textit{R$^2$} value of at least 0.9 and have considered only those values in our analysis.

Among the VOXY monolayers, VOBrI and VOClBr have the highest and lowest values for $e_{11}$ of 13.97 $\times$ 10$^{-10}$ C/m and 11.46 $\times$ 10$^{-10}$ C/m, respectively. The $e_{11}$ value for VOFCl is 11.56 $\times$ 10$^{-10}$ C/m, close to that of VOClBr. In the case of $e_{12}$ piezoelectric coefficient, VOFCl and VOBrI have the highest and lowest values of -3.32  $\times$ 10$^{-10}$ C/m and -2.54 $\times$ 10$^{-10}$ C/m, respectively. VOFCl, VOFBr, and VOFI show significant values for both out-of-plane piezoelectric coefficients, $e_{31}$ and $e_{32}$, while VOClI only has a significant value for $e_{31}$ coefficient. The values for $e_{31}$ coefficient are -0.16, -0.24, -0.42, and -0.19 in units of 10$^{-10}$ C/m for VOFCl, VOFBr, VOFI, and VOClI monolayers, respectively. On the other hand, the values for the $e_{32}$ coefficient are -0.19, -0.25, and -0.25 in units of 10$^{-10}$ C/m for VOFCl, VOFBr, and VOFI monolayers, respectively. Note that the negative sign for piezoelectric coefficients means the reduction of polarization with the applied strain. Smaller values for $e_{12}$ than $e_{11}$ suggest stronger control of ferroelectric polarization via strain engineering along the polar axis, which has also been explored in our previous work on VOCl$_2$ monolayer\cite{Mahajan2021}.
 
The $e_{11}$ values for the VOXY monolayers are an order of magnitude larger than that for both the measured\cite{Zhu2015} (2.9 $\times$ 10$^{-10}$ C/m) and calculated\cite{Duerloo2012}(3.64 $\times$ 10$^{-10}$ C/m) values for MoS$_2$ monolayer, which is a well known 2D piezoelectric material. Although the $e_{12}$ coefficients are much smaller than that for $e_{11}$ coefficients for the VOXY monolayers, they are still comparable to the $e_{11}$ coefficient for the MoS$_2$ monolayer and higher than the $e_{11}$ values calculated for other 2D materials like h-BN\cite{Duerloo2012}(1.38 $\times$ 10$^{-10}$ C/m) and group-III monochalcogenides\cite{Li2015}. The giant piezoelectric coefficient $e_{11}$ values for the VOXY monolayers are comparable to that for the group-IV monochalcogenide monolayers\cite{Fei2015}, that also show the similar anisotropic in-plane piezoelectric effect as shown by VOXY monolayers. The out-of-plane piezoelectric coefficients, $e_{31}$ and $e_{32}$, are of the same magnitude as that of ferromagnetic Janus VSSe monolayer\cite{Zhang2019} (0.948 $\times$ 10$^{-10}$ C/m) and comparable to that for Janus group-III chalcogenide monolayers\cite{Guo2017}, which have values ranging from 0.08 $\times$ 10$^{-10}$ C/m to 0.30 $\times$ 10$^{-10}$ C/m.
                                       
We also calculated the piezoelectric coefficients for VOX$_2$ monolayers, as shown in Table \ref{table5}. Like VOXY monolayers, VOX$_2$ monolayers also have an asymmetric piezoelectric effect with giant values for $e_{11}$ and smaller values with a negative sign for $e_{12}$ piezoelectric coefficients. For VOCl$_2$, VOBr$_2$, and VOI$_2$ monolayers, the $e_{11}$ coefficients are 11.12, 11.92, and 16.90, respectively, in the unit of 10$^{-10}$ C/m, indicating an increase in the piezoelectric effect for strain along \textit{a}-axis with heavier halogen (or larger halogen radii). VOF$_2$ stands as an outlier to the observed trend with $e_{11}$ coefficient value of 12.04 $\times$ 10$^{-10}$ C/m, which is larger than that for VOCl$_2$ and VOBr$_2$ monolayers. VOI$_2$ monolayer has the highest value for $e_{11}$ piezoelectric coefficients compared to all the VOX$_2$ monolayers and all the VOXY monolayers as well. The trend observed in VOX$_2$ monolayers for the increase of piezoelectric effect for strain along \textit{a}-axis with heavier halogen (or larger halogen radii) could explain the trend of $e_{11}$ values in VOXY monolayers, except for some outliers like higher $e_{11}$ values for VOFCl and VOFBr compared to VOClBr, or the higher $e_{11}$ value for VOFI compared to VOClI, that can be explained as a result of VOF$_2$ being an outlier to this trend. The $e_{12}$ coefficient values for VOF$_2$, VOCl$_2$, VOBr$_2$, VOI$_2$ monolayers are -3.91, -2.95, -2.70, and -2.77, respectively, in the unit of 10$^{-10}$ C/m, suggesting a decrease in the piezoelectric effect for strain along \textit{b}-axis with heavier halogen (or larger halogen radii). A similar trend of lower $e_{12}$ values with heavier halogen (or larger halogen radii) is also observed in the VOXY monolayers, leading to the highest $e_{12}$ value for the VOF$_2$ monolayer among all the VOXY and VOX$_2$ monolayers.

\section{Conclusions}
\label{secconclusion}
In summary, we have predicted a new series of 2D Janus multiferroic materials with promising applications in future nanotechnologies, which requires multifunctional materials for multipurpose nanodevices. Our strategic comparison of the properties of the Janus VOXY monolayers with their parent VOX$_2$ monolayers provided better insight into the trend observed in those properties. Here, we would like to emphasize our findings for the VOFCl monolayer, which predicts it to possess collinear ferromagnetism with in-plane ferroelectricity and both in-plane and out-of-plane piezoelectricity. Since experimental data are available for the bulk VOCl$_2$, we encourage experimental studies to exfoliate and obtain VOCl$_2$ monolayer and create a Janus VOFCl monolayer by replacing Cl with F atoms using CVD, a process that we expect thermodynamically more favourable because of higher electronegativity of F compared to Cl. If experimentally realized, multifunctional materials like VOFCl and other Janus VOXY monolayers could accelerate the development of energy-conserving and waste-reducing nanoelectronics and nanosensors. Meanwhile, we also encourage more theoretical studies on these predicted Janus multiferroic VOXY monolayers to enhance our understanding of these systems.                          

\subsection*{Acknowledgements}
We acknowledge funding from SERB (CRG/2021/003687). We also thank computer center IIT Kanpur for providing HPC facility. 
\bibliography{ref}

\begin{table*}[h!]
  \begin{center}
    \caption{Energy of different magnetic configurations (FM, AFM1, AFM2, and AFM3) in meV/f.u.(formula unit) to determine ground-state magnetic configuration (GMC), V ion's magnetic moment (M$_V$) in units of $\mu_B$ (Bohr magneton), and electronic band gap for spin-up and spin-down bands for the Janus VOXY and VOX$_2$ monolayers. Here, E$_{FM}$ is taken as the reference for energy comparison of different magnetic configurations. The asterisk (*) denotes the direct band gap. }
    \label{table1}
    \sisetup{
 	 round-mode          = places, 
  	 round-precision     = 3, 
	}
    \begin{tabular*}{\textwidth}{@{\extracolsep{\fill}} |c c c S S S S S S|}
    \hline
    \hline
      \textbf{Monolayer} & \textbf{GMC} & \textbf{E$\mathbf{_{FM}}$} & \textbf{E$\mathbf{_{AFM1}}$} & \textbf{E$\mathbf{_{AFM2}}$} & \textbf{E$\mathbf{_{AFM3}}$} & \textbf{M$\mathbf{_V}$} & \textbf{Band gap} & \textbf{Band gap}\\
       &   &  &  &  &  &  & \textbf{spin-up}  & \textbf{spin-down}\\  
        &  & \textbf{(meV/f.u.)} & \textbf{(meV/f.u.)} & \textbf{(meV/f.u.)} & \textbf{(meV/f.u.)} & \textbf{($\mu_B$)} & \textbf{(eV)} & \textbf{(eV)}\\
      \hline
      \textbf{VOFCl} & FM & 0 & 5.496 & 3.170 & 5.978 & 1.036 & 0.802 & 2.738* \\
      \hline
      \textbf{VOFBr} & FM & 0 & 7.640 & 4.389 & 9.269 & 1.041 & 0.646 & 1.849* \\
      \hline
      \textbf{VOFI} & FM & 0 & 3.058 & 6.147 & 6.705 & 1.061 & 0.347 & 0.848* \\
      \hline
      \textbf{VOClBr} & AFM3 & 0 & -5.077 & 2.829 & -5.619 & 1.001 & 0.997 & 0.997 \\
      \hline 
      \textbf{VOClI} & FM & 0 & 4.877 & 4.285 & 6.118 & 1.086 & 0.423* & 0.853* \\
      \hline 
      \textbf{VOBrI} & FM & 0 & 1.558 & 4.369 & 2.916 & 1.104 & 0.358* & 0.833* \\
      \hline 
      \textbf{VOF$\mathbf{_2}$} & FM & 0 & 6.568 & 1.279 & 5.968 & 1.037 & 0.884 & 3.513* \\
      \hline 
      \textbf{VOCl$\mathbf{_2}$} & AFM3 & 0 & -11.138 & 2.295 & -11.972 & 0.987 & 1.042 & 1.042 \\
      \hline 
      \textbf{VOBr$\mathbf{_2}$} & AFM3 & 0 & -4.751 & 3.119 & -5.135 & 1.009 & 0.974 & 0.974 \\
      \hline 
      \textbf{VOI$\mathbf{_2}$} & FM & 0 & 1.940 & 3.776 & 3.920 & 1.147 & 0.001* & 0.606* \\
      \hline
      \hline 
    \end{tabular*}
  \end{center}
\end{table*}

\begin{table*}[h!]
  \begin{center}
    \caption{Values for the magnetic exchange coupling parameters and transition temperatures (T$_C$ and T$_N$) for the VOXY and VOX$_2$ monolayers. \textit{J$_a$} and \textit{J$_b$} are nearest-neighbor (NN) exchnage parameters along the \textit{a} and \textit{b} axis, respectively. \textit{J$_{ab}$} is the next-nearest-neighbor (NNN) exchange parameter.Magnetic ordering for each monolayer is also provided in the first column.}
    \label{table2}
    \sisetup{
 	 round-mode          = places, 
  	 round-precision     = 2, 
	}
    \begin{tabular*}{\textwidth}{@{\extracolsep{\fill}} |c S c S S| }
    \hline
    \hline
      \textbf{Monolayer} & \textit{J$\mathbf{_a}$} & \textit{J$\mathbf{_b}$} & \textit{J$\mathbf{_{ab}}$} & $\mathbf{T_C/T_N}$ \\  
        & \textbf{(meV)} & \textbf{(meV)} & \textbf{(meV)} & \textbf{(K)} \\
      \hline
      \textbf{VOFCl (FM)} & 0.86 & 2.00 & 0.31 & 86.69 \\
      \hline
      \textbf{VOFBr (FM)} & 1.41 & 2.98 & 0.30 & 125.50 \\
      \hline
      \textbf{VOFI (FM)} & 2.27 & 0.81 & 0.26 & 94.06 \\
      \hline
      \textbf{VOClBr (AFM3)} & 0.52 & -3.19 & 0.40 & 43.49 \\
      \hline 
      \textbf{VOClI (FM)} & 1.22 & 1.50 & 0.30 & 90.88 \\
      \hline 
      \textbf{VOBrI (FM)} & 1.24 & -0.003 & 0.31 & 52.53 \\
      \hline 
      \textbf{VOF$\mathbf{_2}$ (FM)} & 0.08 & 2.75 & 0.20 & 80.61 \\
      \hline 
      \textbf{VOCl$\mathbf{_2}$ (AFM3)} & 0.35 & -6.11 & 0.38 & 113.05 \\
      \hline 
      \textbf{VOBr$\mathbf{_2}$ (AFM3)} & 0.61 & -2.96 & 0.41 & 36.15\\
      \hline 
      \textbf{VOI$\mathbf{_2}$ (FM)} & 0.39 & 0.52 & 0.13 & 35.23 \\
      \hline
      \hline 
    \end{tabular*}
  \end{center}
\end{table*}

\begin{table*}[h!]
  \begin{center}
    \caption{Magnetic anisotropy energy (MAE) in the units of $\mu$eV/f.u. (formula unit) for the Janus VOXY and VOX$_2$ monolayers. Magnetic ordering for each monolayer is also provided in the first column.}
    \label{table3}
    \sisetup{
 	 round-mode          = places, 
  	 round-precision     = 2, 
	}
    \begin{tabular*}{\textwidth}{@{\extracolsep{\fill}} |c S S S S|}
    \hline
    \hline
      \textbf{Monolayer} & $\mathbf{MAE_{100-010}}$ & $\mathbf{MAE_{001-010}}$ & $\mathbf{MAE_{110-010}}$ & $\mathbf{MAE_{111-010}}$ \\
       & \textbf{($\mathbf{\mu}$eV/f.u.)}& \textbf{($\mathbf{\mu}$eV/f.u.)} & \textbf{($\mathbf{\mu}$eV/f.u.)} & \textbf{($\mathbf{\mu}$eV/f.u.)} \\  
      \hline
      \textbf{VOFCl (FM)} & 9.75 & 4.25 & 5.00 & 2.25 \\
      \hline
      \textbf{VOBrI (FM)} & 40.50 & 173.25 & 20.25 & 72.25 \\
      \hline
      \textbf{VOF$\mathbf{_2}$ (FM)} & 12.50 & 17.50 & 5.00 & 10.00 \\
      \hline
      \textbf{VOI$\mathbf{_2}$ (FM)} & 207.75 & 334.50 & 103.00 & 165.75 \\
      \hline 
      \hline
       & & & & \\
      \hline
      \hline
      \textbf{Monolayer} & $\mathbf{MAE_{010-100}}$ & $\mathbf{MAE_{001-100}}$ & $\mathbf{MAE_{110-100}}$  & $\mathbf{MAE_{111-100}}$ \\  
       & \textbf{($\mathbf{\mu}$eV/f.u.)}& \textbf{($\mathbf{\mu}$eV/f.u.)} & \textbf{($\mathbf{\mu}$eV/f.u.)} & \textbf{($\mathbf{\mu}$eV/f.u.)} \\  
      \hline 
      \textbf{VOFI (FM)} & 81.00 & 12.25 & 40.50 & 57.00 \\
      \hline 
      \textbf{VOClI (FM)} & 15.25 & 67.75 & 7.50 & 16.75 \\
      \hline
      \hline 
       & & & & \\
      \hline
      \hline
      \textbf{Monolayer} & $\mathbf{MAE_{100-001}}$ & $\mathbf{MAE_{010-001}}$ & $\mathbf{MAE_{110-001}}$ & $\mathbf{MAE_{111-001}}$ \\
       & \textbf{($\mathbf{\mu}$eV/f.u.)}& \textbf{($\mathbf{\mu}$eV/f.u.)} & \textbf{($\mathbf{\mu}$eV/f.u.)} & \textbf{($\mathbf{\mu}$eV/f.u.)} \\
      \hline 
      \textbf{VOClBr (AFM3)} & 20.25 & 5.25 & 12.75 & 10.25 \\
      \hline 
      \textbf{VOFBr (FM)} & 58.5 & 50.75 & 54.75 & 27.75 \\
      \hline 
      \textbf{VOCl$\mathbf{_2}$ (AFM3)} & 17.75 & 2.25 & 10.00 & 6.75 \\
      \hline 
      \textbf{VOBr$\mathbf{_2}$ (AFM3)} & 21.00 & 5.00 & 13.00 & 8.75 \\
      \hline 
      \hline 
    \end{tabular*}
  \end{center}
\end{table*}

\begin{table*}[h!]
  \begin{center}
    \caption{Spontaneous in-plane FE polarization (P$_S$); energy barrier for in-plane polarization switching via paraelectric (PE) phase (given as E$_G$), and via antiferroelectric (AFE) phase (given as $\Delta$E), energy difference between FE and AFE phase (brackets shows value for AFE with relaxed lattice parameters), and out-of-plane polarization (P$_{out}$) for Janus VOXY monolayer and VOX$_2$ monolayers.}
    \label{table4}
    \sisetup{
 	 round-mode          = places, 
  	 round-precision     = 2, 
	}
    \begin{tabular*}{\textwidth}{@{\extracolsep{\fill}} |c S S S c c| }
    \hline
    \hline
      \textbf{Monolayer} & $\mathbf{P_S}$ & $\mathbf{E_G}$ & $\mathbf{\Delta E}$ & $\mathbf{FE-AFE}$ & $\mathbf{P_{out}}$ \\  
        & \textbf{($\mathbf{10^{-10}}$ C/m)} & \textbf{(meV/f.u.)} & \textbf{(meV/f.u.)} & \textbf{(meV/f.u.)} & \textbf{($\mathbf{10^{-10}}$ C/m)} \\
      \hline
      \textbf{VOFCl} & 3.03 & 292.11 & 150.22 & 9.75 (10.94) & 0.04 \\
      \hline
      \textbf{VOFBr} & 2.86 & 242.51 & 128.93 & 4.47 (5.07) & 0.07 \\
      \hline
      \textbf{VOFI} & 2.63 & 163.83 & 88.76 & -2.47 (-2.08) & 0.11 \\
      \hline
      \textbf{VOClBr} & 2.70 & 247.47 & 124.58 & -1.32 (-1.24) & 0.04 \\
      \hline 
      \textbf{VOClI} & 2.49 & 173.58 & 82.43 & -1.99 (-1.89) & 0.08 \\
      \hline 
      \textbf{VOBrI} & 2.37 & 145.51 & 71.12 & -0.35 (-0.29) & 0.05 \\
      \hline 
      \textbf{VOF$\mathbf{_2}$} & 3.31 & 338.57 & 161.74 & 28.86 (34.25) & - - - - \\
      \hline 
      \textbf{VOCl$\mathbf{_2}$} & 2.90 & 301.35 & 146.24 & -0.61 (-0.46) & - - - - \\
      \hline 
      \textbf{VOBr$\mathbf{_2}$} & 2.61 & 224.86 & 107.86 & -1.51 (-1.48) & - - - - \\
      \hline 
      \textbf{VOI$\mathbf{_2}$} & 2.13 & 66.00 & 40.97 & 0.73 (1.63) & - - - - \\
      \hline
      \hline 
    \end{tabular*}
  \end{center}
\end{table*}   

\begin{table*}[h!]
  \begin{center}
    \caption{Piezoelectric coefficients for Janus VOXY and VOX$_2$ monolayers.}
    \label{table5}
    \sisetup{
 	 round-mode          = places, 
  	 round-precision     = 2, 
	}
    \begin{tabular*}{\textwidth}{@{\extracolsep{\fill}} |c S c S c| }
    \hline
    \hline
      \textbf{Monolayer} & $\mathbf{e_{11}}$ & $\mathbf{e_{31}}$ & $\mathbf{e_{12}}$ & $\mathbf{e_{32}}$ \\  
        & \textbf{($\mathbf{10^{-10}}$ C/m)} & \textbf{($\mathbf{10^{-10}}$ C/m)} & \textbf{($\mathbf{10^{-10}}$ C/m)} & \textbf{($\mathbf{10^{-10}}$ C/m)} \\
      \hline
      \textbf{VOFCl} & 11.56 & -0.16 & -3.32 & -0.19 \\
      \hline
      \textbf{VOFBr} & 11.93 & -0.24 & -3.09 & -0.25 \\
      \hline
      \textbf{VOFI} & 13.21 & -0.42 & -2.90 & -0.25 \\
      \hline
      \textbf{VOClBr} & 11.46 & - - - - & -2.89 & - - - - \\
      \hline 
      \textbf{VOClI} & 13.04 & -0.19 & -2.62 & - - - - \\
      \hline 
      \textbf{VOBrI} & 13.97 & - - - - & -2.54 & - - - - \\
      \hline 
      \textbf{VOF$\mathbf{_2}$} & 12.04 & - - - - & -3.91 & - - - - \\
      \hline 
      \textbf{VOCl$\mathbf{_2}$} & 11.12 & - - - - & -2.95 & - - - - \\
      \hline 
      \textbf{VOBr$\mathbf{_2}$} & 11.92 & - - - - & -2.70 & - - - - \\
      \hline 
      \textbf{VOI$\mathbf{_2}$} & 16.90 & - - - - & -2.77 & - - - - \\
      \hline
      \hline 
    \end{tabular*}
  \end{center}
\end{table*}

\end{document}